\documentclass[prl,aps,twocolumn,showpacs]{revtex4-1}
\usepackage{graphicx}
\usepackage{dcolumn}
\usepackage{bm}
\usepackage{color}
\usepackage{amssymb,amsmath}

\begin{document}

\title{Universal monomer dynamics of a two dimensional 
semi-flexible chain}

\author{Aiqun Huang}
\affiliation{Department of Physics, University of Central Florida, Orlando, Florida
32816-2385, USA}
\author{Ramesh Adhikari}
\affiliation{Department of Physics, University of Central Florida, Orlando, Florida
32816-2385, USA}
\author{Aniket Bhattacharya}
\altaffiliation[]
{Author to whom the correspondence should be addressed}
\email{aniket@physics.ucf.edu}
\affiliation{Department of Physics, University of Central Florida, Orlando, Florida
32816-2385, USA}
\author{Kurt Binder}
\affiliation{Institut f\"ur Physik, Johannes Gutenberg-Universit\"at Mainz, 
Staudinger Weg 7,55099, Mainz, Germany}

\date{August 9, 2013}

\begin{abstract}
We present a unified scaling theory for the dynamics of monomers for dilute solutions of semiflexible
polymers under good solvent conditions in the free draining limit. Our theory encompasses the well-known
regimes of mean square displacements (MSDs) of stiff chains growing like $t^{3/4}$ with time due to bending motions,
and the Rouse-like regime $t^{2 \nu / (1+ 2\nu)}$ where $\nu$ is the Flory exponent describing the radius $R$ of
a swollen  flexible coil. We identify how the prefactors of these laws scale with the persistence length $\ell_p$, and
show that a crossover from stiff to flexible behavior occurs at a MSD of order $\ell^2_p$ (at a time proportional to $\ell^3_p$).
A second crossover (to diffusive motion) occurs when the MSD is of order $R^2$. Large scale Molecular Dynamics
simulations of a bead-spring model with a bond bending potential (allowing to vary $\ell_p$ from 1 to 200
Lennard-Jones units) provide compelling evidence for the theory, in $D=2$ dimensions where $\nu=3/4$.
Our results should be valuable for understanding the dynamics of DNA (and other semiflexible biopolymers)
adsorbed on substrates. 
\end{abstract}
\maketitle
Conformations and dynamics of semi-flexible polymers in bulk as well as under various applied fields, {\em e.g.}, confining 
and stretching potentials are of broad general interest in different disciplines of science. Important 
biopolymers, {\em e.g.}, dsDNA, F-Actin, microtubules, all have finite 
bending rigidity $\kappa$, often with large persistence lengths and hence the well established and matured theories for fully flexible
chains often are not adequate to describe these biopolymers as flexural rigidity 
plays an important role in their mechanical responses~\cite{Physical_Biology}. 
Interests in these biopolymers continue to remain unabated for multiple reasons.
(i) A deeper understanding of biopolymers, {\em e.g.}, Actin, Titin, Fibrin 
which offer intriguing patterns with unusual viscoelastic responses will allow to design bio-mimetic materials with improved 
characteristics, not seen in synthetic polymers; (ii) there is a genuine need to develop efficient separation methods 
of biomolecules, {\em e.g.}, DNA sequencing and separation of proteins for various applications pertaining to health and diseases.
Finally, due to advent of sophisticated single molecule probes, {\em e.g.}, fluorescence correlation spectroscopy, 
atomic force microscope spectroscopy, scanning electron spectroscopy with which one can directly observe the dynamics of the 
entire chain as well as fluorescence labeled segments of these biomolecules~\cite{Sackmann_NJP}-\cite{Rechendorff} 
which offer new findings to be further explored.\par
Historically the worm-like chain (WLC) model has been the paradigm for theoretical studies of semi-flexible chains. The 
Hamiltonian for the WLC is given by
\begin{equation}       
\mathcal H = \frac{\kappa}{2} \int_0^L \left (\frac{\partial^2 \mathbf{r}}{\partial s^2}\right)^2ds,
\label{wlc}
\end{equation}
where $L$ is the contour length, $\kappa$ is the bending rigidity and the integration is carried along the contour 
$s$~\cite{Rubinstein,Doi}.
One can show that in 2D and 3D dimensions $\ell_p = 2\kappa/k_BT$ and $\kappa/k_BT$ respectively~\cite{Landau}. The model has been 
studied quite extensively applying path integral and other techniques~\cite{Yamakawa}-\cite{Wilhelm_PRL_1996} and exact expressions 
of various moments of the distribution of monomer distances along the chain have been worked out.
In particular, the end-to-end distance for the WLC model is given by~\cite{Rubinstein}
\begin{equation}
\frac{\langle R_N^2\rangle}{L^2} = \frac{2}{n_P}\left(1-\frac{1}{n_p}[1-\exp(-n_p)]\right),
\label{rn_wlc}
\end{equation}
where $L= (N-1)\delta$ is the contour length and $n_p = L/\ell_p$. Here we recall that any linear polymer is a chain molecule of 
$N$ discrete monomeric units, and we take the distance between the neighboring units as $\delta$. 
In the limit $n_P >> 1$, {\em i.e.}, $\ell_p << L$ one gets $\langle R_N^2\rangle = 2\ell_pL$ and the chain behaves like a Gaussian coil;  
for  $n_P << 1$, $\langle R_N^2\rangle = L^2$ and the chain behaves like a rod.
Evidently the model neglects the excluded volume (EV) interaction and hence interpolates between rod and Gaussian limit. 
Dynamics of the WLC model has been explored using Langevin type of equation~\cite{Granek_1997,Maggs_MM_1993,Kroy_PRE_1997,Wilhelm_PRL_1996,Bullerjahn_EPL_2011}. 
One can expect that the dynamics of a stiff-chain will be dominated by transverse fluctuations (bending modes)~\cite{Winkler_JCP_2003} and that  
the short time dynamics will be governed by the chain persistence length. Indeed a relaxation dynamics using the WLC 
Hamiltonian (Eqn.~\ref{wlc}) approach yields an expression for fluctuation 
$\langle \left( \Delta h\right)^2  \rangle \sim \ell_p^{-0.25}t^{0.75}$, 
which crosses over to simple diffusion at late time~\cite{Granek_1997,Maggs_MM_1993}. This $t^{0.75}$ behavior has been observed in 
many experiments using flouroscence probe and video microscopy on F-Actin network~\cite{Sackmann_NJP,Goff_PRL_2002,Caspi_PRL_1998} and in 
some simulations of polymer network~\cite{Bulacu,Steinhauser}.
Analytical studies of monomer dynamics in a WLC model, similar to \cite{Granek_1997,Maggs_MM_1993} have been carried 
out for a tagged particle by Bullerjahn {\em et al.}~\cite{Bullerjahn_EPL_2011} who also found that the transverse MSD of a tagged particle obeys subdiffusive behavior of $t^{0.75}$. \par
While these predictions based on WLC model are consistent with some of the experiments, the WLC model fails to capture important 
aspects caused by EV effects \cite{Hsu_EPL_2010,Hsu_EPL_2011} invalidating Eqn.~\ref{rn_wlc} in the limit $n_p >> 1$ both in 2D and 3D where 
the chain statistics in $D$ spatial dimension satisfies~\cite{Pincus_MM_1980,Nakanishi_1987}, 
\begin{equation}
\sqrt{\langle R_N^2 \rangle} \sim N^{\nu}\ell_p^{{1}/{D+2}}.
\label{saw}
\end{equation}
The Gaussian regime of WLC model is completely absent in 2D~\cite{Hsu_EPL_2011}; in 3D the 
Gaussian regime crosses over at $\langle R_N^2 \rangle \sim \ell_p^3$ to 3D self avoiding walk (SAW) of Eqn.~\ref{saw}~\cite{Hsu_EPL_2010}. 
Furthermore, the angular correlation between subsequent bonds along the chain, instead of exponential, as predicted by the WLC model, 
exhibits a power-law decay. Therefore, EV effect has a profound effect on the statistics of stiff chains as well.\par
A key question is then how the EV effect 
affects the monomer dynamics of a semiflexible chain. We have developed a scaling theory of monomer dynamics for a compressible semi-flexible 
chain. {\em We predict a novel double crossover dynamics where the initial sub-diffusive relaxation of the monomers characterized by a $t^{0.75}$ law 
at an intermediate time crosses over to the monomer dynamics of a flexible chain $t^{\frac{2\nu}{1+2\nu}}$ before reaching 
the purely diffusive regime for the entire chain.} This is the main theoretical result of this letter. 
We support our claim by carrying out extensive 
BD simulation for a large number of chain lengths from $N=16$ to $N=1024$ and $\kappa=1.0 - 128$, 
to show that (i) $\langle R_N^2\rangle/(2L\ell_p)$ as a function of $L/\ell_p$ for all ratios $L/\ell_p$ collapse on the same master plot 
and that the early time slope of unity ($\langle R_N^2\rangle \propto L^2$; rod limit) directly crosses over to slope of 0.5
($\langle R_N^2\rangle \propto L^{1.5}\ell_p^{0.5}$; 2D SAW, Eqn.~\ref{saw}) clearly demonstrating absence of Gaussian regime in 2D.
(ii) Second, by monitoring the dynamics of middle monomer $g_1(t) = \langle \left(\mathbf{r}_{N/2}(t) - \mathbf{r}_{N/2}(0)\right)^2\rangle $, and comparing it 
with that of the center of mass $g_3(t) = \langle \left(\mathbf{r}_{CM}(t) - \mathbf{r}_{CM}(0)\right)^2\rangle $, and the relative dynamics of $g_1(t)$
with respect to $\mathbf{r}_{CM}(t)$ expressed as 
$g_2(t) =  \langle \left( \mathbf{r}_{N/2}(t)- \mathbf{r}_{CM}(t)) - (\mathbf{r}_{N/2}(0)- \mathbf{r}_{CM}(0) \right)^2 \rangle$
~\cite{Grest_Kremer_PRA_1986,Gerroff_JCP_1992,Milchev_JCP_1993,Binder_Review} we show data
collapse and monomer crossover dynamics. We believe these studies of chain conformation and monomer dynamics 
will be extremely valuable to interpret experimental results and testing certain approximations in analytical theories 
for semiflexible chains~\cite{Kroy_PRE_1997,Wilhelm_PRL_1996,Harnau1}. \par 
$\bullet$~{\em Scaling theory}:~
We start with the Eqn.~\ref{Granek} below derived by Granek and Maggs~\cite{Granek_1997,Maggs_MM_1993}  using a Langevin dynamics framework for  
the WLC Hamiltonian of Eqn.~\ref{wlc}  
\begin{equation}
g_1(t) = \delta^2\left(\delta/\ell_p\right)^{1/4}\left(Wt\right)^{3/4}.
\label{Granek}
\end{equation}
Here we have chosen the inverse of a monomer reorientation rate $W^{-1}$ as the unit of time. 
For early time the monomer dynamics will be independent of the chain length $N$
until the fluctuations become of the order of $\ell_p$. Therefore, denoting the first crossover occurs at time $\tau_1$ and substituting 
$g_1 = \ell_p^2$ and $t= \tau_1$ in Eqn.~\ref{Granek} we immediately get  
\begin{equation}
W\tau_1 = \left(\ell_p/\delta\right)^3.
\label{tau1}
\end{equation}
For $ 0 < t \le W^{-1}(\ell_p/\delta)^3$  the monomer dynamics is described by $g_1(t) \sim t^{0.75}$ until 
$g_1(t) = \ell_p^2$ at time  $W^{-1}(\ell_p/\delta)^3$.  
The width of this region is independent of $N$ and solely a function of $\ell_p$. \par
\begin{figure}[ht!]                
\centering
\includegraphics[width=0.8\columnwidth]{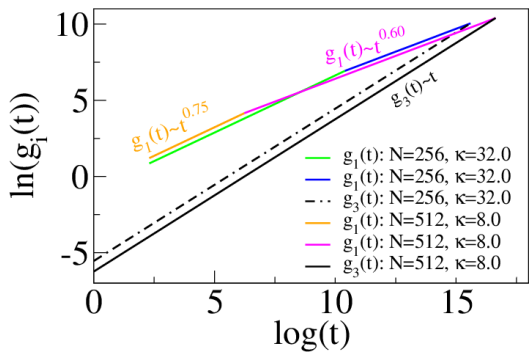}\\
\caption{\small Theoretical scaling plots for $(N,\kappa)\equiv (256,32)$  and 
$(N,\kappa)\equiv (512,8)$. Green and orange lines correspond to $g_1(t) \sim t^{0.75}$, blue and magenta lines correspond to $g_1 \sim t^{0.60}$,
and the dashed and solid black lines correspond to $g_3(t) \sim t$ for $N=256$ and 512 respectively. 
The width of each region shows how these regimes depend on $\ell_p$ and $N$. Note that in reality we expect a very 
gradual change of slope on the log-log plot at both crossover times, rather than sharp kinks.}
\label{phase}
\end{figure}

 
For $ \tau_1 < t < \tau_2$ the dynamics is governed by the Rouse relaxation of monomers of a fully flexible EV chain in 2D characterized by 
$g_1(t) = t^{2\nu/(1+2\nu)} = t^{0.6}$. $\tau_2$ characterizes the onset of  
the purely diffusive regime when $g_1(\tau_2) = \langle R_N^2 \rangle$~\cite{Grest_Kremer_PRA_1986}.   
We then obtain $\tau_2$ as follows:
\begin{equation}
g_1(t) = \ell_p^2 \left( t/\tau_1 \right) ^{3/5} \mathrm{~~~~~~for~~} t > \tau_1.
\end{equation}
Substituting $\tau_1$ from Eqn.~\ref{tau1} in above
\begin{equation}
g_1(t)  =  \delta^2 \left(\ell_p/\delta\right)^{1/5} \left( Wt\right)^{3/5},  \mathrm{~~for~~}  \tau_1 < t <  \tau_2.
\label{gmid}
\end{equation}
At $t = \tau_2$ 
\begin{equation}
g_1(t=\tau_2)  =   \langle R_N^2 \rangle = \ell_p^{1/2} \delta^{3/2}N^{3/2}.
\end{equation}
Substituting Eqn.~\ref{gmid} for $t=\tau_2$ we get
\begin{equation}
W \tau_2  =   \left(\ell_p/\delta\right)^{\frac{1}{2}}N^{5/2}.
\label{tau2}
\end{equation}
We also note that the dynamics of the center of mass is given by (omitting prefactors of order unity throughout) 
\begin{equation}
g_3(t) = \delta^2 W \frac{t}{N}.
\end{equation}
The ``phase diagram'' for the crossover dynamics in terms of $N$, and $\ell_p$ are shown in Fig.~\ref{phase}. 
Notice that for a stiffer chain the 
region for $\tau_1 < t < \tau_2$ for which we predict $g_1(t) \sim t^{0.6}$ becomes progressively small and therefore, is  
hard to see in simulation for a stiffer chain. \par
$\bullet$~{\em BD simulation for a 2D EV semiflexible chain}: 
We have used a standard BD scheme  using 
Lennard-Jones(LJ), finite extensible nonlinear elastic (FENE)~\cite{Grest_Kremer_PRA_1986}  potentials to describe 
EV and spring potentials, and a three body potential among three consecutive monomers to describe chain stiffness as in 
\cite{Adhikari_JCP_2013,Supplementary}. \par
\begin{figure}[ht!]                
\begin{center}
\includegraphics[width=0.9\columnwidth]{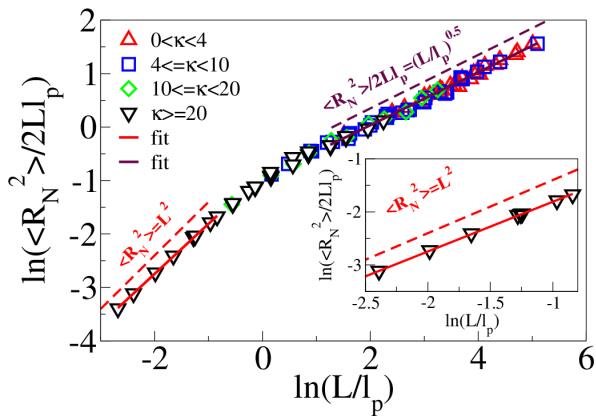}
\end{center}
\vskip -0.5truecm
\caption{\small $\langle \langle R_N^2\rangle/(2L\ell_p)$ as a function of $L/\ell_p$ obtained from 
different combination of chain length $N$ and stiffness parameter $\kappa$ (log-log scale). The solid (maroon) line 
is a fit to the formula $\langle R_N^2\rangle/2L\ell_p \sim (L/\ell_p)^{0.5}$ for $4 < L/\ell_p < 160$.
The inset shows the same for small values of $0 < L/\ell_p< 1$ which clearly indicates that limiting 
slope of unity ($\langle R_N^2 \rangle = L^2$) for $L/\ell_p \rightarrow 0$. }
\label{rn}
\end{figure}

$\bullet$~{\em Absence of Gaussian regime and correct scaling for a 2D EV chain:}~
Fig.~\ref{rn} shows a plot of $\langle R_N^2\rangle/2\ell_pL$  as a function of $L/\ell_p $  for a huge number of values of $L/\ell_p$ ($\sim 100$).
For $L/\ell_p << 1$ we observe that $\frac{\langle R_N^2\rangle}{2\ell_pL} \sim (L/\ell_p)^{1.0}$ while for 
$L/\ell_p >> 1$ the data very nicely fit with $\frac{\langle R_N^2\rangle}{2\ell_pL} \sim (L/\ell_p)^{0.50}$.
This plot for chains with varying degree of stiffness and chain length conclusively shows the absence of Gaussian regime in a 2D EV chain earlier 
observed by Hsu {\em et al.} from a lattice model~\cite{Hsu_EPL_2011} and observed in experiments 
with single stranded DNA on modified graphite substrate~\cite{Rechendorff}.\par
\begin{figure}[ht!]                
\centering
\includegraphics[width=\columnwidth]{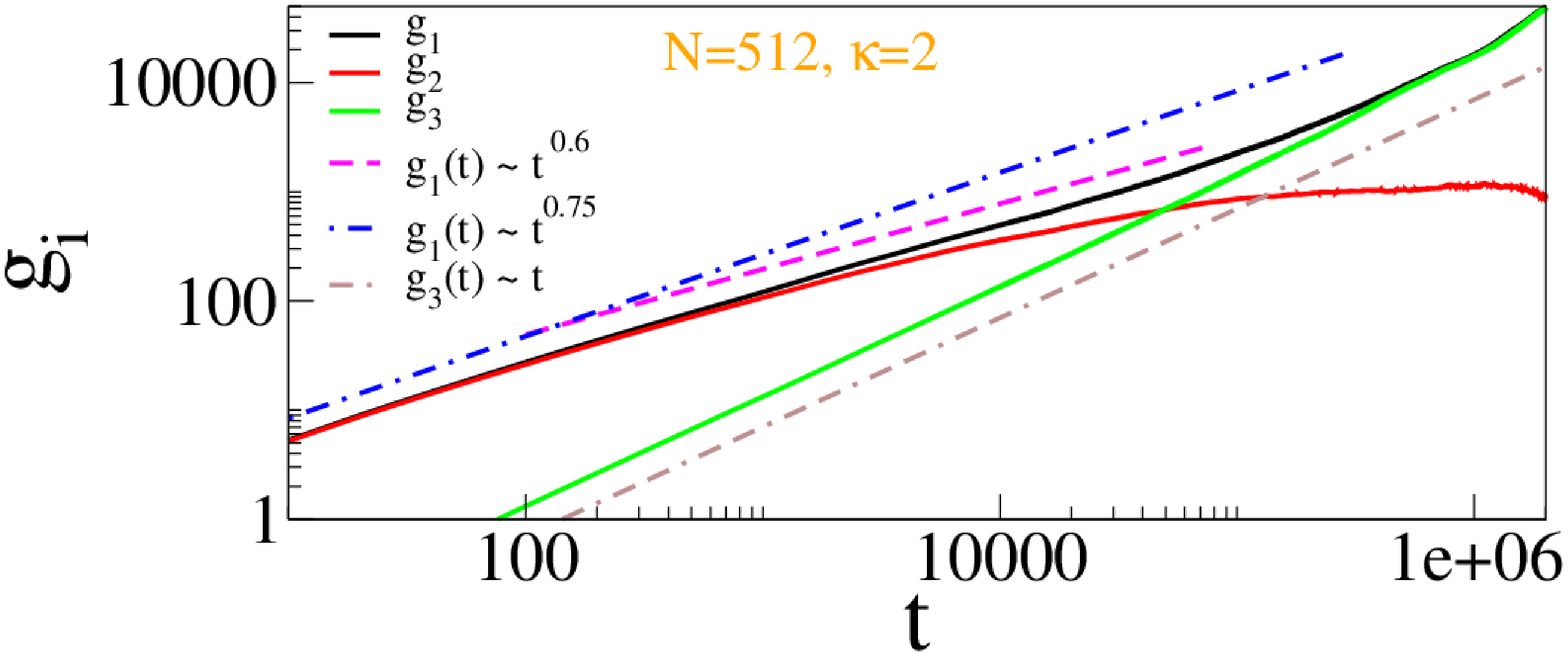}\\
\includegraphics[width=\columnwidth]{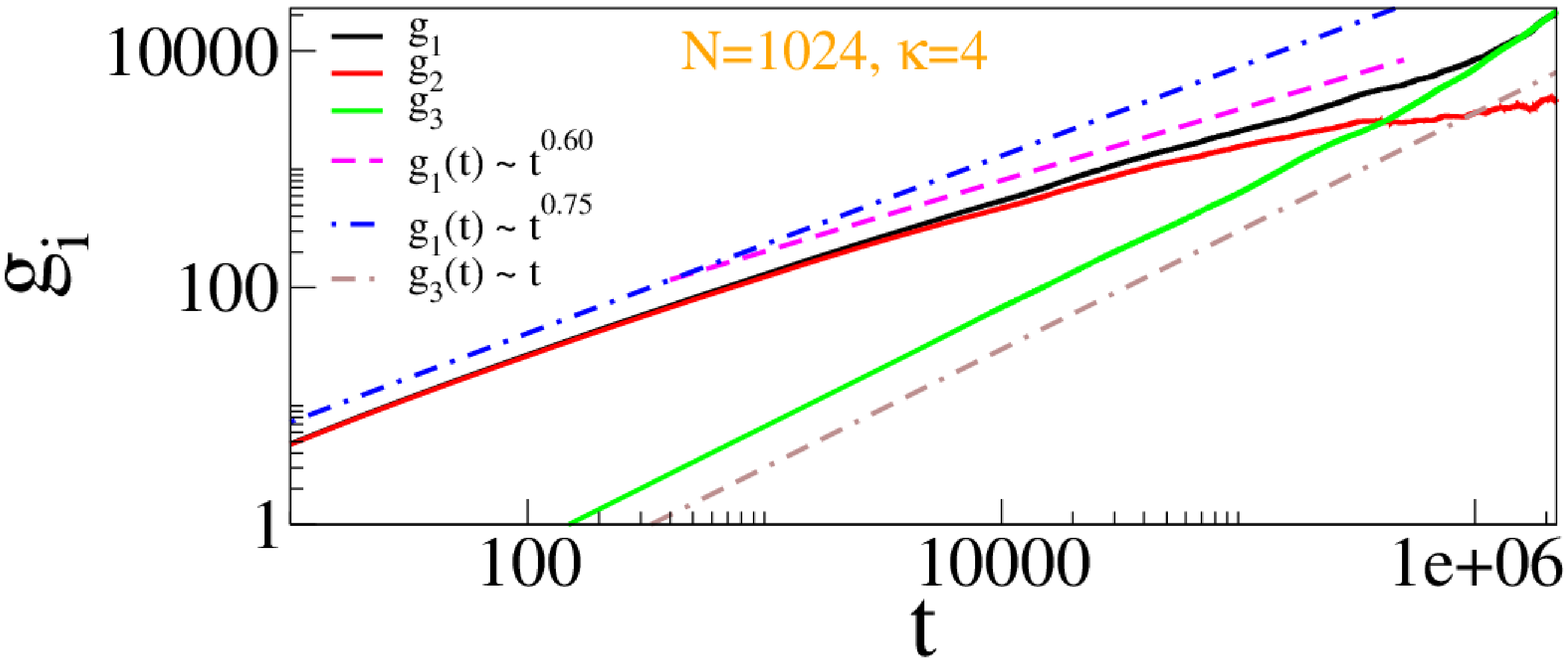}\\
\caption{\small (top) Plot for $g_1(t)$ (black), $g_2(t)$ (red) and $g_3(t)$ (green) as a function of time on a log-log scale for 
chain length $N=512$ and $\kappa = 2.0$.
The blue and magenta dashed lines correspond to straight lines $g_1(t) = At^{0.75}$,  and $g_1(t) = Bt^{0.60}$, respectively, 
where $A$ and $B$ are constants. (bottom) same but for $N=1024$ and $\kappa = 4.0$.} 
\label{n512}
\end{figure}
\begin{figure}[ht!]                
\centering
\includegraphics[width=1.0\columnwidth]{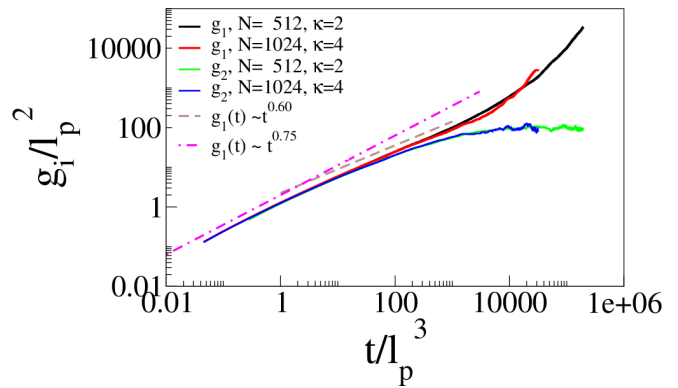}
\caption{\small Plot for $g_1(t)/\ell_p^2$ (black and red)  and $g_2(t)/\ell_p^2$ (blue and green)
as a function of $t/\ell_p^3$ on a log-log scale for chain lengths $N = 512$, $\kappa=2.0$ 
and for $N = 1024$, $\kappa = 4.0$ respectively. The dot-dashed lines correspond to slopes 0.75 (magenta) and 0.6 (brown) respectively.}
\label{g1g2}
\end{figure}

$\bullet$~{\em Monomer dynamics:}~We now present BD simulation results to confirm our scaling theory.  
Results for $g_1(t)$, $g_2(t)$, and $g_3(t)$  shown in Fig.~\ref{n512} unambiguously confirm our predictions. These plots 
quite clearly show three distinct scaling regimes of $g_1(t) \sim t^{0.75}$ crossing over to $g_1(t) \sim t^{0.6}$ and then merging with $g_3(t) \sim t$ at late times.
The double crossover required simulation of reasonably large chain lengths ($N=512-1024$) below 
which it is hard to see these crossovers conclusively. 
Fig.~\ref{g1g2} shows plot of $ g_1(t)/\ell_p^2$ as a function of rescaled time $t/\ell_p^3 $ which shows data collapse for various chain length $N$ and $\kappa$ again 
confirming the time scales for these crossovers. As expected, the crossovers are rather gradual, spread out over a decade in time $t$ 
each, and hence for chains that are not long enough the existence of these regimes is easily missed. \par
To summarize, we have provided a new scaling theory of monomer dynamics for semiflexible polymers in 2D. Our theory predicts novel crossover dynamics at an intermediate 
time when the fluctuations of the monomers become greater than $\ell_p$. Around this time the monomer dynamics becomes the same as that of a fully flexible 
chain characterized by $g_1(t) \sim t^{2\nu/(1+2\nu)} =  t^{0.6}$ in 2D . The theory expands the existing scaling 
theory for monomer dynamics for a WLC and that of a fully flexible chain to include the effect of the chain persistence length. 
Fully flexible chains are self-similar objects, while a polymer segment up to its own 
persistence length is not. Therefore, it is expected that for length scale up to $\ell_p$ the dynamics will have different characteristics due to bending modes arising 
out of the chain stiffness. The EV effect is almost negligible for the $t^{0.75}$ regime and therefore, our result is the same as that of from previous studies using 
WLC Hamiltonian~\cite{Granek_1997,Maggs_MM_1993}. For the $t^{0.6}$ regime originating from EV effect, 
where the monomer dynamics is governed by Rouse relaxation of a fully flexible chain, 
our theory elucidates the exact role of chain persistence length neither contained in WLC model nor seen before. 
We also validate our new scaling theory by extensive BD simulation results. 
\par
We now comment on generalization of our results in 3D and/or in presence of hydrodynamic(HD) interactions. In the free draining limit 
the $t^{0.75}$ regime will remain the same in 3D~\cite{Granek_1997,Maggs_MM_1993}, but the intermediate Rouse relaxation regime will be characterized by 
$t^{2\nu/(1+2\nu)} = t^{0.54}$ ($\nu = 0.59$ in 3D). Replacing Rouse relaxation by Zimm relaxation one immediately 
sees that in presence of HD interaction the intermediate regime is characterized by $\sim t^{2\nu/3\nu}=t^{2/3}$~\cite{Hinczewski}. 
Notice that in this case $\nu$ cancels out and this relaxation should be the same in 2D and 3D.  
Our results are completely consistent with the observed $t^{0.5}$ and $t^{2/3}$ power laws due to Rouse and Zimm relaxation for the monomers in 
double and single stranded DNA using flouroscence correlation spectroscopy by Shusterman {\em et al.}~\cite{Shusterman_PRL_2004}.\par
Finally, we provide plausible explanation why this double crossover has not been seen in 
single molecule experiments with biopolymers~\cite{Sackmann_NJP,Goff_PRL_2002, Caspi_PRL_1998,Shusterman_PRL_2004}. 
A simple calculation for Fig.~\ref{phase} shows that in order for the width of the $t^{0.75}$ and $t^{0.60}$ to 
be equal (in logarithmic scale) one needs $N = l_p^{2.2}$ in 2D. In other words for a stiffer chain one needs 
a very long chain to see the $t^{0.60}$ regime. Indeed in our simulation we found 
(not shown here) that for $\kappa = 16$, 32, and 64, the results with chain length up to $N=512$ are largely dominated by 
the $t^{0.75}$ regime and we did 
not clearly see the  $t^{0.60}$ regime. It is only after we lowered the value of $\kappa$ and used longer chain ($N=1024$), we identified these 
two regimes quite conclusively (Fig.~\ref{n512}).
We suspect that the same might happen in experiments~\cite{Sackmann_NJP}. For extreme stiff chains the $t^{0.6}$ (or $t^{0.54}$ in 3D) region can be extremely narrow 
and could either be easily missed or the rather smooth double crossover might be mistakenly interpreted as a 
single crossover (with $t^{2/3}$ in 2D). 
Therefore, we believe that these results will not only promote new experiments but will be extremely valuable in identifying and 
interpreting different scaling regimes for the monomer dynamics of semiflexible polymers.\par
AB, AH, and RA acknowledge financial support through a seed grant from UCF.

\end{document}


\title{Universal monomer dynamics of a two dimensional 
semi-flexible chain: Supplemental Materials}
\author{Aiqun Huang}
\affiliation{Department of Physics, University of Central Florida, Orlando, Florida
32816-2385, USA}
\author{Ramesh Adhikari}
\affiliation{Department of Physics, University of Central Florida, Orlando, Florida
32816-2385, USA}
\author{Aniket Bhattacharya}
\affiliation{Department of Physics, University of Central Florida, Orlando, Florida
32816-2385, USA}
\author{Kurt Binder}
\affiliation{Institut f\"ur Physik, Johannes Gutenberg-Universit\"at Mainz, 
Staudinger Weg 7,55099, Mainz, Germany}
\date{Aug 09, 2013}
\maketitle
\vskip -20 truecm
\section{The model and  details of Brownian Dynamics simulation}
\begin{figure}[ht!]                
\centering
\includegraphics[width=0.5\columnwidth]{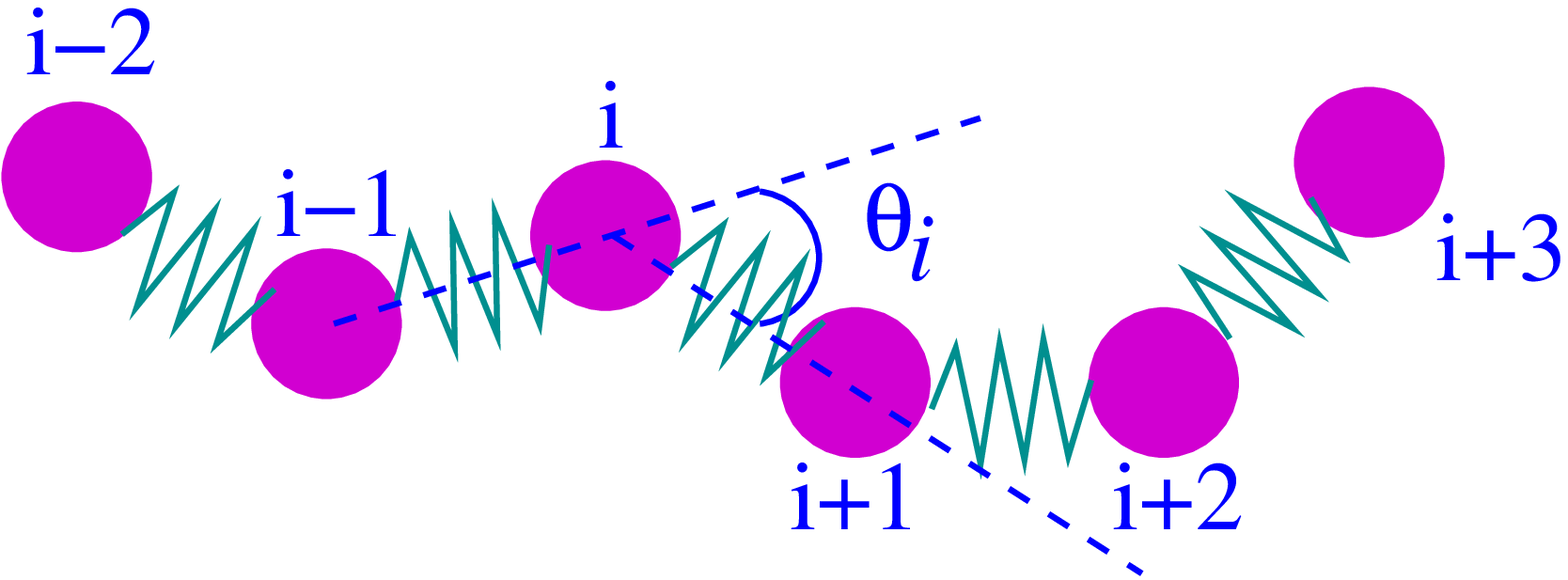}
\caption{\small Bead-spring model of a polymer chain with bending 
angle $\theta_i$ subtended by the vectors 
$\vec{b}_{i} = \vec{r}_{i}-\vec{r}_{i-1}$ and  $\vec{b}_{i+1} = \vec{r}_{i+1}-\vec{r}_{i}$.}
\label{model1}
\end{figure}

We have carried out Brownian dynamics (BD) simulation using a a bead spring model for a semiflexible polymer chain (Fig.~\ref{model1}) 
with excluded volume (EV), spring and 
bond bending potentials as described below~\cite{Grest}.
The excluded volume interaction between any two monomers is given by short range Lennard-Jones (LJ) potential.
\begin{eqnarray}
U_{\mathrm{LJ}}(r)&=&4\epsilon [{(\frac{\sigma}{r})}^{12}-{(\frac{\sigma}
{r})}^6]+\epsilon \;\mathrm{for~~} r\le 2^{1/6}\sigma \nonumber \\
        &=& 0 \;\mathrm{for~~} r >  2^{1/6}\sigma\;.
\end{eqnarray}
Here, $\sigma$ is the effective diameter of a monomer, and
$\epsilon$ is the strength of the potential. The connectivity between
neighboring monomers is modeled as a Finite Extensible Nonlinear
Elastic (FENE) spring with 
\begin{equation}
U_{\mathrm{FENE}}(r)=-\frac{1}{2}kR_0^2\ln(1-r^2/R_0^2)\;, 
\end{equation}
where $r$ is the distance
between consecutive monomers, $k$ is the spring constant, and $R_0$
is the maximum allowed separation between connected monomers~\cite{Grest}.
The chain stiffness is introduced by adding an angle dependent interaction between successive bonds (discrete worm-like-chain potential)
as 
\begin{equation}
U_{\mathrm{BEND}}(\theta_i) = \kappa(1-\cos \theta_i).
\label{kappa}
\end{equation}
Here $\theta_i$ is the angle between the bond vectors 
$\vec{b}_{i-1} = \vec{r}_{i}-\vec{r}_{i-1}$ and 
 $\vec{b}_{i} = \vec{r}_{i+1}-\vec{r}_{i}$, respectively, as shown in Fig.~\ref{model1}. The strength 
of the interaction is characterized by the bending rigidity $\kappa$.\par
We use the Brownian dynamics with the following equation of motion for the i$^{th}$ monomer 
\begin{equation}
m \ddot{\vec{r}}_i=-\vec{\nabla} (U_\mathrm{LJ}+U_\mathrm{FENE} + U_\mathrm{BEND})-\zeta \vec{v}_i 
+ \vec{\eta}_i.
\end{equation}
Here $\zeta$ is the monomer friction coefficient and
$\vec{\eta} _ i (t)$, is a Gaussian white noise with zero mean at a temperature T, and
satisfies the fluctuation-dissipation relation: 
\begin{equation} 
< \, \vec{\eta} _ i (t)
\cdot \vec{\eta} _ j (t') \, > = 4k_BT \zeta \, \delta _{ij} \, \delta (t
- t ')\;.
\end{equation}
The reduced units of length, time and temperature are chosen to be  $\sigma$, 
$\sigma\sqrt{\frac{m}{\epsilon}}$, and $\epsilon/k_B$ respectively.  
For the spring potential we have chosen $k=30$ and $R_0=1.5\sigma$, the friction coefficient 
$\zeta = 0.7$, and the temperature $T=1.2/k_B$. 
The choice of the FENE potential along with the LJ interaction parameters 
ensures that the average bond-length in the bulk $\langle b_l \rangle = 0.971$. With the choice of these 
parameters the probability of chain crossing is very low. 
The LJ and the FENE parameters are chosen to be the same as in our previous studies
of polymer translocation of semiflexible chains~\cite{Adhikari_JCP_2013}. 
We advance the position coordinates using a reduced time step $\Delta t = 0.01$ following the 
algorithm proposed by van Gunsteren and Berendsen~\cite{Langevin}.\par
We have carried out simulation for  
various combinations of $N$ (16 - 1024) and $\kappa$ (1 - 100) and sampled a wide range of values of $n_p = L/\ell_p$ (please 
see Eqn. 2 in the main text) to simulate semiflexible chains with varying degree of stiffness. 
With these choices of parameters we find that the average bond-length $\langle b_l \rangle$
is hardly affected by the chain stiffness parameter. During the simulation, first we have equilibrated each 
chain for 5-100 times of the Rouse relaxation time, and then collected data over 
30-250 times of the Rouse relaxation time in order to obtain good statistics for $g_1(t)$, $g_2(t)$, and $g_3(t)$. 
To calculate crossover dynamics and the exponents it took a 16-core Intel processor almost a month to finish the most 
time-consuming job ($N$=1024).\par
\begin{figure}[ht!]                
\begin{center}
\includegraphics[width=0.42\columnwidth]{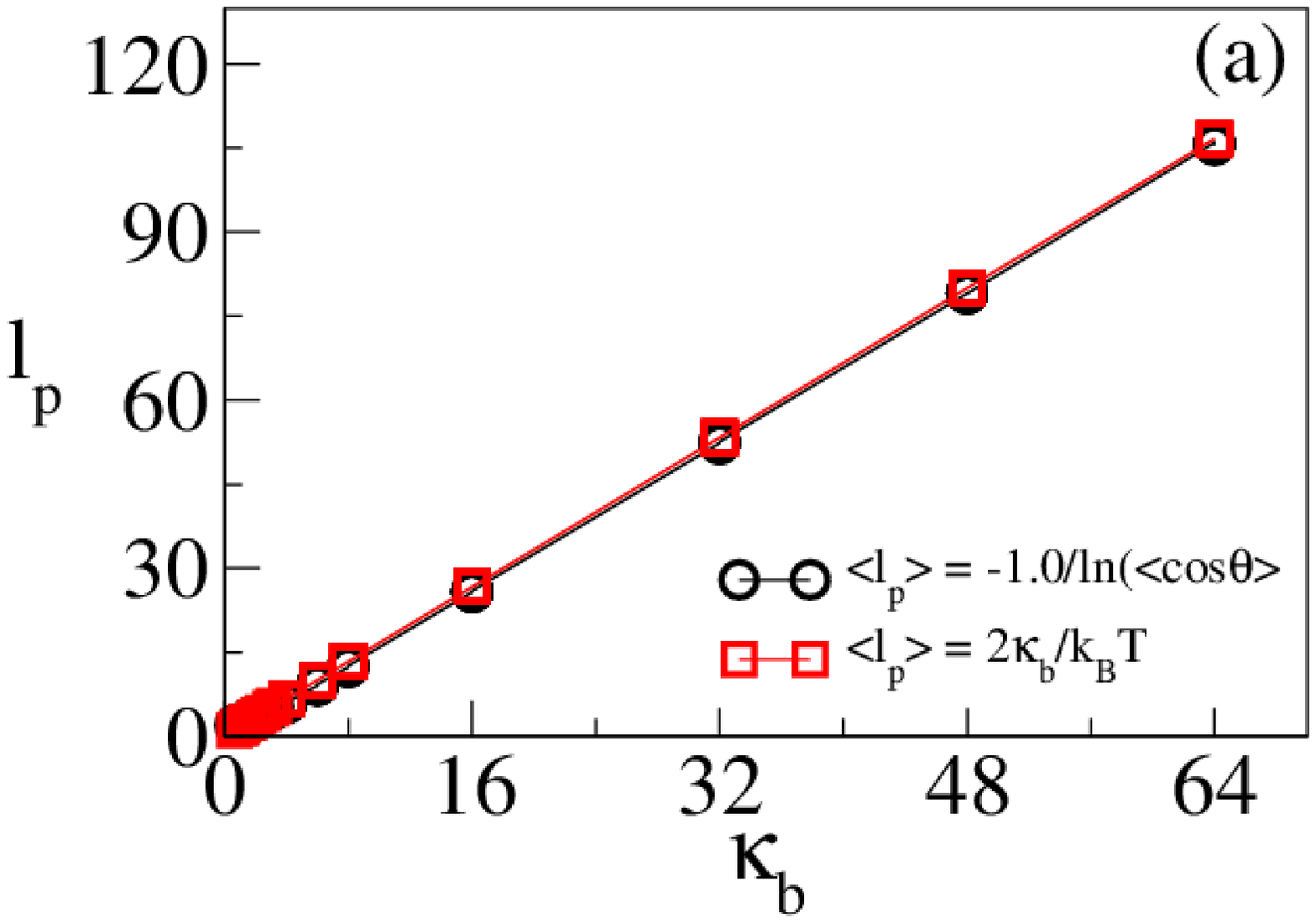}
\includegraphics[width=0.45\columnwidth]{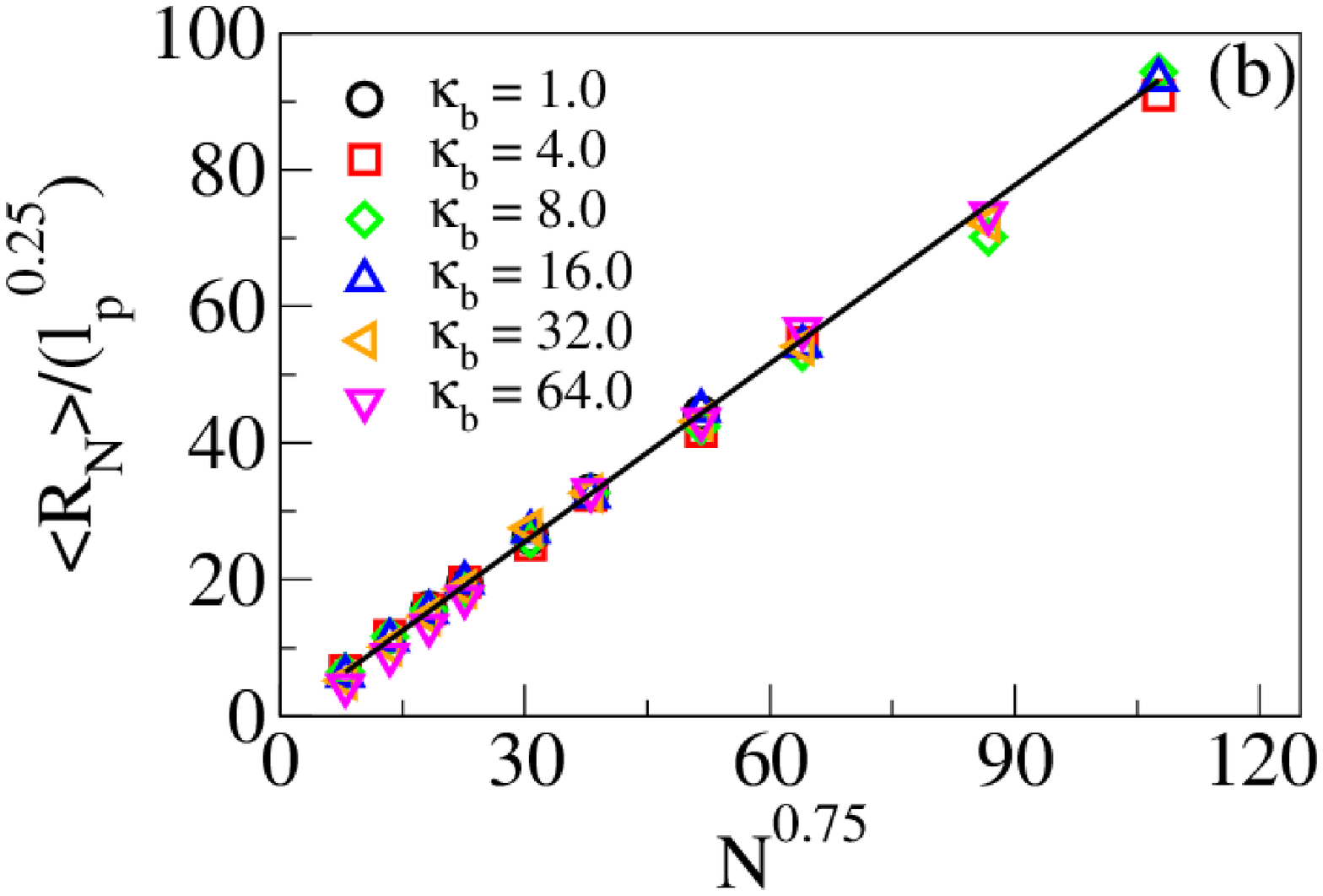}
\end{center}
\caption{\small (a) Comparison of $\ell_p = -1/\ln(\langle \cos \theta \rangle)$ and $\ell_p = 2\kappa/k_BT$. 
(b) Plot of $\sqrt{\langle R_N^2 \rangle}/\ell_p^{0.25}$ versus $N^{0.75}$ for 
various values of the chain stiffness parameter. All the data for different 
stiffness parameters collapse on the same master plot. The solid line is a fit to a straight line}. 
\label{lp}
\end{figure}

Using the BD algorithm we first calculated the chain persistence length  $\ell_p$ as a function of the bending 
rigidity $\kappa$.  It is worth noting that the persistence 
length must not be extracted from the decay of bond orientational correlations at large distances along the chain, 
but even in presence of the EV interaction it still can be estimated from the average $\langle \cos\theta \rangle$ between subsequent bonds by the 
standard formula $\ell_p = -1/\ln \langle \cos \theta \rangle $. 
We find that this estimate also coincides with the continuum theory result $\ell_p=2\kappa/k_BT$~\cite{Rubinstein} as shown 
in Fig.~\ref{lp}(a) where we have used 
$\langle \cos \theta \rangle$ obtained from simulation to calculate  $\ell_p$ for different combination of $\kappa$ and $N$. 
Since persistence length is an intrinsic local property of the chain, EV has very little effect on it. 
This although not the main focus of this letter is a new result~\cite{comment1}. Using $\ell_p$ obtained from 
simulation we then verified that in 2D the end-to-end distance satisfies the relation 
$\sqrt{\langle R_N^2 \rangle} \sim N^{\nu}\ell_p^{0.25}$ ~\cite{Pincus_MM_1980,Nakanishi_1987}
for various combinations of $\ell_p$ and chain length $N$~\cite{Adhikari_JCP_2013} as shown in Fig~\ref{lp}(b). 
We have used these values of $\ell_p$ and $\sqrt{\langle R_N^2 \rangle}$ in the main text. \par
